\documentclass[review,3p,11pt]{article}

\usepackage{graphicx}
\usepackage{algorithmic}
\usepackage{bbm}
\usepackage{xcolor}
\usepackage{amsfonts,dsfont}
\usepackage{multirow}
\usepackage{mathtools}
\usepackage{amsmath,amsfonts,amssymb}
\usepackage{booktabs}
\usepackage{float}
\usepackage{multicol}
\usepackage{svg}
\DeclareMathOperator*{\argmin}{arg\,min}

\begin{document}

\title{Sparse2Inverse: Self-supervised inversion of
sparse-view CT data}

\author{{{Nadja Gruber \thanks{Corresponding author: nadja.gruber@uibk.ac.at}} $^,$\thanks{Department of Mathematics, University of Innsbruck, Austria}} $^,$\thanks{VASCage-Research Centre on Vascular Ageing and Stroke, Innsbruck, Austria} 
\and
Johannes Schwab \thanks{MRC Laboratory of Molecular Biology, Cambridge, UK}
\and Elke Gizewski \thanks{Department of Neuroradiology, Medical University of Innsbruck, Austria}
\and Markus Haltmeier \footnotemark[2]}

\maketitle

\begin{abstract}
Sparse-view computed tomography (CT) enables fast and low-dose CT imaging, an essential feature for patient-save medical imaging and rapid non-destructive testing. In sparse-view CT, only a few projection views are acquired, causing standard  reconstructions to suffer from severe artifacts and noise. To address these issues,  we propose a self-supervised image reconstruction strategy. Specifically, in contrast to the established Noise2Inverse, our proposed training strategy uses a loss function in the projection domain, thereby bypassing the otherwise prescribed nullspace component. We demonstrate the effectiveness of the proposed method in reducing stripe-artifacts and noise, even from highly sparse data.
\end{abstract}

%

\section{Introduction}



Image reconstruction from sparse CT measurements suffers from data incompleteness and a large null space, resulting in severe non-uniqueness. In addition, the inversion is unstable, causing noise in the data to be amplified during the reconstruction process. These issues result in reconstructed images containing severe streak-artifacts and noise. Although conventional filtered back projection (FBP) is fast and accurate for full data, it produces low-quality images dominated by noise and artifacts in sparse-view CT. This highlights the importance of regularization that incorporates specific prior knowledge for high-quality reconstruction. Recently, machine learning methods have significantly improved reconstruction quality and have gained considerable popularity in this context \cite{li2020nett,arridge2019solving,schwab2019deep,romano2017little}.

Classical variational  and iterative reconstruction methods allow for the efficient incorporation of hand-crafted priors such as smoothness or total variation \cite{scherzer2009variational}. While accurate in many scenarios, they rely on idealized prior assumptions, are time consuming, and are highly sensitive to the choice of regularization parameters. Recently, deep learning methods have gained popularity because they naturally adapt to the available image structure and produce high quality reconstructions. However, these methods typically require supervised training data with specific input-target pairs tailored to a particular tomographic setting.




Self-supervised learning combines the advantages of both iterative and learning-based methods and has recently become an important direction in deep learning and image processing. It overcomes the need for data annotation while exploiting the power of learnable neural networks. Self-supervised methods developed for denoising tasks exploit inherent structure and redundancy in the data to remove noise from signals and images. Self-supervised methods can also be extended to reconstruct missing or corrupted parts of a signal or image, where the goal is to fill in missing information or to restore the original signal from noisy or incomplete observations.
Existing studies on self-supervised CT image reconstruction operate either entirely in the image domain \cite{hendriksen2020noise2inverse} or entirely in the data domain \cite{zhou2022low, zhang2021noise2context}. Methods that operate only in the image domain do not perform well for operators with a large null space, such as sparse-view CT. On the other hand, methods that operate only in the data domain may not perform optimally because convolutional neural networks are built for image structure rather than projection data.

In this work, we propose a self-supervised learning strategy for image reconstruction in sparse-view CT.  Our method allows the CNN to operate in the image domain and combine it with a loss in the data domain.  In contrast to Noise2Inverse, our method can handle a large null space, allowing the simultaneous removal of noise and streak-artifacts while maintaining data consistency.

\section{Related Work}

\subsection{Self-supervise denoising}

In~\cite{lehtinen2018noise2noise}, the authors  introduce Noise2Noise, a self-supervised  denoising methods using only noisy image pairs $(x_t \odot \delta_{t,1}, x_t \odot \delta_{t,2})$ for $i=1, \dots, N$. They showed that training solely with corrupted images, assuming independent and additive noise, achieves results comparable to noisy-clean pairs.
However, obtaining multiple instances of a signal with  different  noise  may also be challenging in practice.
This limitation was eliminated with Noise2Self~\cite{batson2019noise2self}, which enables self-supervised training with only noisy measurements.

In Noise2Self, pixels from the input image are (randomly) masked out and the neural network is trained to predict values of the masked pixels from the noisy unmasked pixels.
Training a neural network using the Noise2Self strategy approximates training with ground truth targets if the noise has mean zero and is element-wise independent across different pixels in the image.


\subsection{Self-supervised image  reconstruction}

Training without the requirement of clean targets is also valuable for image reconstruction  problems of the  form $y = A x \odot  \delta$ such as CT or MRI reconstruction. In this section we review existing approaches that are relevant for our  strategy.

\paragraph{Image and projection domain methods}

Natural self-supervised strategies for image reconstruction are to apply Noise2Self either in the image domain or in the data domain.  In image domain methods, a reconstruction operator $B$ such as Filtered Backprojection (FBP) is applied in a first step, and then Noise2Self is applied, treating $BAx$ as the ground truth and $B \delta$ as the reconstructed noise. However, $B\delta$ is statistically coupled, which severely degrades Noise2Self. Another option is to use self-supervised training schemes in the data domain ~\cite{zhou2022low,zhang2021noise2context} where the noise is statistically independent. Since neural networks tend to perform better on real images than on sinograms, this degrades performance. In addition, both methods are unable to handle image components orthogonal to the nullspace of $A$.

\paragraph{Noise2Inverse}

Some drawbacks of image and projection domain methods are addressed by Noise2Inverse~\cite{hendriksen2020noise2inverse}, which extends Noise2Self to image reconstruction in the case of complete data. Here, the data is split in the projection domain, where the noise is statistically independent, and a network is trained in the reconstruction domain.
Noise2Inverse aims to remove measurement noise, but is not designed to remove artifacts resulting from under-sampling in the case of sparse data. The reason for this is that the targets in this framework are defined using FBP and thus still fix the nullspace component of the reconstructions.


 \paragraph{Extensions}

In~\cite{chen2021equivariant}, the authors propose a self-supervised learning strategy for the inverse problem, assuming that the underlying signal distribution is invariant to the action of a group of transformations.  In~\cite{unal2021self}, the authors proposed an adaptation of the Noise2Self strategy that operates in the data domain combined with a differentiable learnable backprojection. In contrast to this approach, we adapt the sampling strategy proposed in \cite{hendriksen2020noise2inverse} to the sparse data case combined with a different training loss.


\paragraph{MRI-reconstruction}

Closely related to our work are methods for self-supervised MRI reconstruction from sparsely sampled and noisy data~\cite{yaman2020self, millard2023theoretical, millard2023clean, blumenthal2022nlinv}. However, while the Fourier transform of MRI is an isometry, the Radon transform used in sparse-view CT lacks this property. In addition, the sampling pattern used in these MRI studies is Cartesian (with the exception of~\cite{blumenthal2022nlinv}), whereas the sparse CT data is related to radial MRI data. 


\section{Method}\label{sec:method}

In this section we describe the proposed self-supervised learning strategy for sparse-view CT reconstruction. For the sake of simplicity, we consider the parallel beam geometry in 2D. However, the overall concept can easily be extended to more general geomeries in 2D or 3D.

\subsection{Problem Formulation}

Let $x\in X = \mathbb{R}^{m}$ represent a discretized ground-truth image and  denote by  $y = A_{\Sigma}x  \odot\delta \in Y$ the noisy sparse view CT data at equidistant angles $\Sigma\coloneqq\{\sigma_1, \dots,\sigma_l \}$ on the semicircle corresponding to normal vectors of the measurement lines.
Here $\odot\delta$ denotes impact of measurement noise which might be non-additive.  We consider the situation, where the Radon data are available only for a small number of directions, meaning that $\Sigma$ only has a few elements.

Following the self-supervised learning paradigm we assume that we are given a collection of noisy data
\begin{align}\label{reconproblem}
    y_t^\delta = A_{\Sigma}x_t \odot\delta_t \quad \text{ for } t=1, \dots, N
    \end{align}
without access to the clean and artifact-free ground truth images $x_t \in X$. The aim is to construct a  reconstruction operator $R \colon Y \to X$ using  \eqref{reconproblem}, mapping noisy sparse data to  reconstructions close to unknown ground-truth images.

\subsection{Proposed Sparse2Inverse}
 
\paragraph{Network architecture}

For  given  numbers $k, p\in\mathbb{N}$ we define subsets of projection directions $\Theta_i = \{\sigma_j\mid j \operatorname{mod}k = i\}$, consider the  partition $\boldsymbol{\Theta}\coloneqq\{\Theta_1,\ldots,\Theta_k\}$ and  we use $\mathcal{I}$ to denote the collection of all subsets $I\subset\boldsymbol{\Theta}$ with $p$ elements. Further let $A_\Theta$  and $B_\Theta$ 
denote the Radon transform and corresponding FBP  with directions in $\Theta$ and write  $M_\Theta y$ for the restriction of data  $y$  to directions in $\Theta$. 
Inspired by \cite{hendriksen2020noise2inverse}  we take the desired reconstruction operator as a parameterized network of the form
\begin{align} \label{eq:recon1}
R_{\psi}(y^\delta) &\coloneqq  
\frac{1}{|\mathcal{I}|}
\sum_{I \in \mathcal{I}} 
R_{\psi,I}(y^\delta)
 \\ \label{eq:recon2}
R_{\psi,I}(y^\delta) &\coloneqq
\Phi_{\psi}
\left(
\frac{1}{p}\sum_{\Theta \in I} B_\Theta ( M_\Theta y^\delta ) 
\right) \,,
\end{align} 
where  $\Phi_\psi \colon X\to X$ is a neural network parametrized by $\psi$,
 
The final reconstructions at inference are taken as $R_{\psi^*}(y^\delta)$ where $\psi^*$ optimizes the loss defined next.

\paragraph{Self-supervised loss}

\begin{figure}[H]
    \centering
\includegraphics[width = 0.99\textwidth]{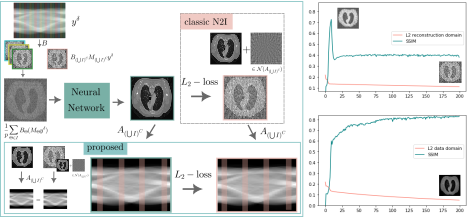}
    \caption{
The proposed training strategy of Sparse2Inverse, in comparison with the classic Noise2Inverse approach~\cite{hendriksen2020noise2inverse}. Both methods start with an initial data splitting in the projection domain, where the noise is pixel-wise independent. The neural network takes the mean sub-reconstruction as input, and the target is defined by the remaining data. Unlike the classic approach, where the loss is computed in the reconstruction domain, our strategy applies the Radon transform and computes the loss in the data domain. The effectiveness of this approach is evident in the curves on the right.}
    \label{scheme}
\end{figure}

The proposed strategy, and its comparison with the classic Noise2Inverse approach  is visualized in Figure~\ref{scheme}. Note that opposed to Noise2Inverse,  we use a loss in the projection domain instead of a loss in the image domain. More precisely, for Sparse2Inverse we construct the network parameters as $\psi^\ast \in \argmin_{\psi} \mathcal{L}_Y(\phi)$ using the loss
\begin{align}\label{loss}
    \mathcal{L}_Y(\phi) 
    = 
    \sum_{t=1}^N \sum_{I\in \mathcal{I}} \lVert A_{(\bigcup I)^C} (R_{\psi,I}(y^\delta_t)) -  M_{(\bigcup I)^C} y^\delta_t  \rVert_2^2.\,
\end{align}
Thus, we apply the neural network in the reconstruction domain, and evaluate the loss  in the projection domain for angles not belonging to any of the sets in $I$ for any partition $I \in  \mathcal{I}$ and  any training data.   
\subsection{Comparison to Noise2Inverse}

Note that due to the large nullspace of the sparse-view forward operator, our loss \eqref{loss} significantly differs from the Noise2Inverse loss 
\begin{align}\label{loss2}
\mathcal{L}_{X}(\phi)
    =
    \sum_{i=1}^N \sum_{I \in \mathcal{I}} \lVert
     R_{\psi,I}(y^\delta_t) -  B_{(\bigcup{I})^C }M_{(\bigcup{I})^C}y^\delta_t  \rVert_2^2 \,.
     \end{align}
This discrepancy can be attributed to the fact that opposed to \eqref{loss}, the loss \eqref{loss2} targets $B_{(\bigcup{I})^C} A_{(\bigcup{I})^C}$ and thus learns streak-artifacts due to $\Sigma$. 
More formally, this behaviour of  Noise2Inverse method can be summarized as follows:
Any ground truth image $x$, can be decomposed as
$x = x^{\ast} + v,$ with $v\in\mathcal{N}(A_{\Theta})$ and $x^\ast \in \mathcal{R}(A_\Theta^T)$. The minimum norm solution of a noise-free equation yields $B_{\Theta}(y) = x^\ast$.
Looking at the Noise2Inverse approach~\cite{hendriksen2020noise2inverse}, where the loss is computed in the reconstuction doamin,
there is no chance for the network to learn the correct nullspace component $v$ as it is fixed to $v = 0$.
 The target therefore specifies a fixed nullspace element, which the neural network $\Phi_\Psi$ aims to learn.

In the Sparse2Inverse approach, despite the neural network being fed images containing artifacts from the FBP, the optimization process of the Convolutional Neural Network (CNN) may  converge to a solution with fewer artifacts that still adequately explains the measured data. Consequently, compared to the reconstruction domain loss, there is no inherent motivation for the network to alter the solution and prefer an alternative one.
\section{Experiments}

The source code, data, and experiments are accessible on GitHub\footnote{https://github.com/Nadja1611/Self-supervised-image-reconstruction-for-Sparse-view-CT.git}. Our proposed method is evaluated on two datasets: a subset of 32 images from the CBCT Walnut dataset (Cone-Beam X-Ray CT Data Collection Designed for Machine Learning)~\cite{der2019cone} and 30 images from the Chest CT-Scan Images Dataset\footnote{https://www.kaggle.com/datasets/mohamedhanyyy/chest-ctscan-images}. We employed the UNet architecture~\cite{ronneberger2015u}, with an image resolution of $336\times 336$, and uniformly distributed projections between 0 and $\pi$. The neural network was trained for 2000 epochs using the Adam optimizer with a learning rate of 0.0002. Our method was compared with Noise2Inverse~\cite{hendriksen2020noise2inverse}, FBP, and total variation minimization (TV). For the implementation of~\cite{hendriksen2020noise2inverse}, we utilized the same UNet architecture and a learning rate of 0.0001.

\begin{table}
   \label{tab:example_multicolumn}
   \small
   \centering
   \begin{tabular}{lcccccc}
   \toprule\toprule
   \multicolumn{7}{c}{Walnuts (Photon count 1000)}  \\

    & \multicolumn{2}{c}{16 angles}&  \multicolumn{2}{c}{32 angles} & \multicolumn{2}{c}{64 angles}       \\
   \midrule
       & SSIM & PSNR &SSIM & PSNR &SSIM &PSNR    \\ \hline
  FBP & 0.100 & 7.743& 0.082 & 9.712& 0.156&18.231\\
  TV & \textbf{0.650}& 17.882& 0.788 & 20.867 &0.838 & 22.606    \\
Noise2Inverse~\cite{hendriksen2020noise2inverse} & 0.408& 15.402 &0.500 & 18.800 &0.736 &21.793  \\
      Proposed & \textbf{0.650} & \textbf{19.103} & \textbf{0.798} & \textbf{22.347} &\textbf{0.857} &\textbf{24.373} \\
   \bottomrule
   \end{tabular}
      \caption{Evaluation metrics SSIM and PSNR of the reconstructions obtained on the Walnut dataset for the different methods.}
\end{table}

\begin{table}
   \label{tablungs}
   \small
   \centering
   \begin{tabular}{lcccccc}
   \toprule\toprule
  \multicolumn{7}{c}{lung CT (Photon count 3000)}    \\
   \multicolumn{2}{c}{16 angles}&  \multicolumn{2}{c}{32 angles} & \multicolumn{2}{c}{64 angles}  \\
   \midrule
       & SSIM & PSNR &SSIM & PSNR &SSIM &PSNR  \\ \hline
  FBP & 0.054&10.574&0.117 & 14.176& 0.220 & 17.552\\
  TV  &\textbf{0.905}& \textbf{29.201} &  0.925 & 30.613 & 0.935 & 31.627   \\
Noise2Inverse~\cite{hendriksen2020noise2inverse}  & 0.586 & 25.775 &  0.673 & 28.356 & 0.743 &30.663 \\
      Proposed  & 0.860 & 28.088 & \textbf{0.928} &   \textbf{31.394}& \textbf{0.956} & \textbf{34.006} \\
   \bottomrule
   \end{tabular}
      \caption{Evaluation metrics SSIM and PSNR of the reconstructions obtained on the lung CT data for the different methods.}
\end{table}

Projection images underwent Poisson noise corruption with photon counts of 1000 (walnuts) and 3000 (lung CTs), following the methodology outlined in~\cite{hendriksen2020noise2inverse}. The sinograms were subsequently divided into four parts (denoted as $k=4$). For the implementation of FBP and TV, we utilized tomosipo~\cite{hendriksen-2021-tomos} and leveraged LION\footnote{https://github.com/CambridgeCIA/LION/tree/main} for specifying the CT geometry.

\begin{figure}[H]
\center
    \includegraphics[width = 0.91\textwidth]{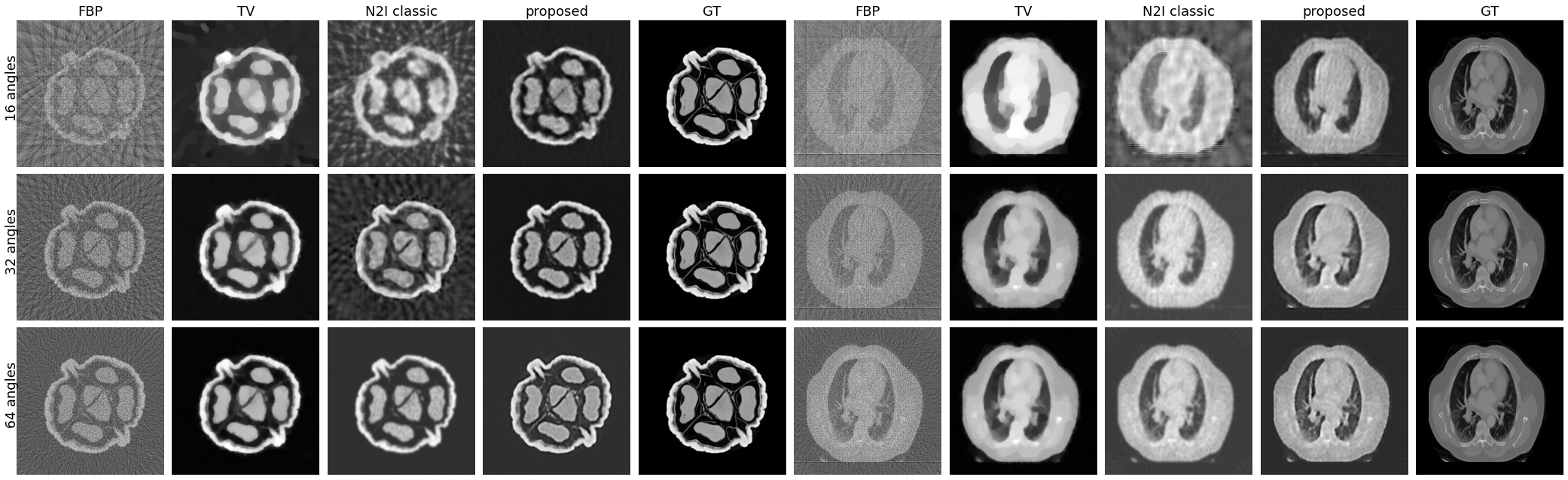}
    \caption{\textbf{Comparison of the reconstructions obtained from the four different methods on the two different datasets for different numbers of projection angles.} We clearly see the streak-artifacts present in the classic Noise2Inverse reconstructions.}\label{results_figure}
\end{figure}

In the assessment and comparison phase, we employed the Structural Similarity Index (SSIM) and Peak Signal to Noise Ratio (PSNR). 
To optimize the total variation (TV) minimization process, we conducted a grid search for the optimal regularization parameters $\lambda$ for SSIM and PSNR. 
During the training of the proposed method and the classic Noise2Inverse approach, evaluation metrics were computed using the ground truth image, and reconstructions were selected to maximize SSIM and PSNR. Although the ground truth was used in this work we made the observation that SSIM and PSNR improve until the end of the training.

We further note that, during training we have observed that the SSIM of the classic Noise2Inverse reaches its maximum at a quite early point of training, and the results at that point have less artifacts than later, when the training loss is minimized. This is also underlined on the right side of Figure~\ref{scheme}.

Figure~\ref{results_figure} shows the reconstructions obtained from the different methods and three sparse view geometries. We observe that compared with the classical Noise2Inverse~\cite{hendriksen2020noise2inverse}, the proposed approach leads to reconstructions with less streak-artifacts and finer details. In addition, the smoothing effect of the results is lower compared to~\cite{hendriksen2020noise2inverse}, and additionally, we observe that for more views also the reconstructions obtained from~\cite{hendriksen2020noise2inverse} have fewer artifacts as the network may have problem in learning high-frequency artifacts.

\section{Conclusion}
In this work, we propose Sparse2Inverse, a self-supervised method for the reconstruction from sparse-view CT data. We present a strategy, where the neural network operates in the reconstruction domain, while the loss is computed in the projection domain, and show its superiority for severely undersampled data in comparison with the classic Noise2Inverse approach~\cite{hendriksen2020noise2inverse} and classical reconstruction algorithms like FBP and TV minimization.

\bibliographystyle{plain}
\bibliography{main}
\end{document}